\documentclass[12pt]{iopart}

\begin{document}

\title{Cooling in the Universe}
\author{Sohrab Rahvar}

\address{Department of Physics, Sharif University of Technology,\\
P.O.Box 11365-9161, Tehran, Iran.\\
and\\
Institute for Studies in Theoretical Physics and Mathematics,
P.O.Box 19395-5531, Tehran, Iran}

\ead{rahvar@sharif.edu}
\begin{abstract}

One of the questions in the cosmology courses is the cooling
mechanism of cosmic fluid during it expansion according to
classical concepts of the thermodynamics. In this short
pedagogical paper, we quote the questions and give a natural
approach dealing with this problem by measuring the dispersion
velocity of the particles in the cosmic fluid by a comoving
observer. We show that the thermal motion of the particles in the
cosmic fluid deviates from the Hubble flow and follows the
geodesics governed by the gravity of the homogeneous universe. The
dynamics of the "thermal peculiar velocity" of the particles leads
in an expanding universe, momentum of the particles relax by the
inverse of the expansion factor and the result is losing the
energy of particles, hence cooling the universe.
\end{abstract}


\maketitle

One of the main questions of students of the cosmology is the
cooling mechanism of the universe during its expansion. According
to the concepts from the classical thermodynamics the question are
as: (i) Is our universe non-adiabatic (i.e. universe is connected
to the cold thermal bath and losses its energy during the time) ?
The answer to this question is trivial, there is nothing outside
the universe, so thermodynamically our universe is adiabatic.(ii)
So, if our universe expands adiabatically the particles of gas
should loos their energy during the expansion of the universe by
bouncing off from the physical boundary of the universe during the
expansion which makes the cosmic fluid to cool down.

In the standard cosmological text books \cite{pady,winberg} and
pedagogical articles \cite{amr1,amr2,amr3,amr4}, the variation of
temperature of the cosmic fluid during the expansion of the
universe has been discussed in several ways. It has been shown
that the momentum of particles in the cosmic fluid decrease with
the inverse of scale factor of the expansion, however some of
these approaches such as dealing the motion of particles with the
special theory of relativity seems does not satisfy the students's
curiosity \cite{pady}. Here we use the natural general
relativistic approach to revisit this problem.

In the standard model of the cosmology the distance between any
two points in an expanding homogeneous universe is given by $r=ax$
where $a$ is the expansion factor and $x$ is defined as the
comoving distance. As the universe expands, the scale factor
increases and makes a larger distance between any two points in
the universe. For an homogeneous universe the comoving distance
does not change with time, so the velocity of expansion can be
given by $v_h = x {da}/{dt}$, so-called {\it Hubble velocity}. For
the real universe, the density is not homogenous and due to the
large structures and voids the density of matter deviates from the
background density of universe. One of the consequences of this
deviation is that there is no more a solid comoving coordinate
system and it changes by time as the structures grow. So the
velocity of cosmic fluid gets one more term of $v_{pec}= a
{dx}/{dt}$, so-called peculiar velocity, due to the variation of
comoving distance with time.

 We take a comoving observer freely follows the expansion of the
universe. For this observer the global velocity of the cosmic
fluid around him/here is zero. However due to none-zero
temperature of the cosmic fluid, the observer will measure
non-zero velocity due to the thermal motion of cosmic fluid, we
call this velocity as {\it thermal peculiar velocity}. We should
bear in mind that thermal peculiar velocity is different from that
appear due to the perturbation in the gravitational potential of
the large scale structures.
The dynamics of the (thermal motion of) particles in the
Friedman-Robertson-Walker ({\it FRW}) universe with the metric of
$ds^2 = dt^2 - a^2dx^2$ is given by the geodesic equation as:
\begin{equation}
{\ddot x^{\mu}} + \Gamma^{\mu}_{\nu \lambda} {\dot x}^{\nu}{\dot x}^{\lambda} = 0,
\label{geodesic}
\end{equation}
where derivation is taken with respect to the comoving time of the
particles, $\tau$ and $\Gamma^{\mu}_{\nu \lambda}$ is the
Christoffel symbol in the ({\it FRW}) metric.
$\Gamma^{i}_{i0}=1/a\times{da/dt}$ is the only non-zero spatial
component of the Christoffel symbol in this metric and by
substituting it in (\ref{geodesic}), the equation of motion of the
particles in the comoving frame obtain as:
\begin{equation}
{\ddot x^{i}}+ \frac{2}{a}\frac{da}{dt}\times \frac{dt}{d\tau} \dot x^i = 0
\end{equation}
The solution of this equation implies: ${dx}/{d \tau}\propto
a^{-2}$. Using the chain rule to express this result in terms of differentiation
with respect to t-component results:
\begin{equation}
\frac{dx}{dt}\frac{dt}{d\tau} \propto a^{-2}.
\label{pec}
\end{equation}
Now we use the definition of peculiar velocity $v_{pec} =
adx/{dt}$ and substitute $dt/{d \tau} = (1-v_{pec}^2)^{-1/2}$ in
equation (\ref{pec}) which yields:
 \begin{equation}
\frac{v_{pec}}{\sqrt{1-v_{pec}^2}}\propto \frac{1}{a}.
\label{pp}
\end{equation}
By multiplying both sides of the equation (\ref{pp}) to the rest
mass of a particle, the momentum of thermal motion of particles obtain as:
\begin{equation}
p = {m_0}v_{pec}/{\sqrt{1-v_{pec}^2}}\propto \frac{1}{a}.
\end{equation}
This equation shows that the momentum of (non)-relativistic
particles relaxes as $1/a$ during the expansion of the universe.
We see that for a static universe, momentum would remain constant.
Decreasing the momentum of the particles during the expansion of
the universe causes the energy of the ultra-relativistic
particles, $E = pc$ to fall as $1/a$. In the case of
non-relativistic particles where $E\propto p^2$, the energy will
fall with a faster rate of $1/a^2$. Using the distribution
function of particles for the (non)-relativistic particles, the
temperature of cosmic fluid scales with the mean energy of the
particles as $T\simeq E$ \cite{pad02}. Now according to the
dependence of the energy of the particles to the scale factor we
can come to conclude that the relativistic particles will cool
with the rate of 1/a while the non-relativistic particles will
cool with a faster rate of
$1/a^2$.\\

Summarizing this letter, we assigned the dispersion velocity of
the particles of the cosmic fluid to the peculiar velocity in a
comoving reference frame and obtained the dependence of the
momentum of particles to the scale factor both for the
(non)-relativistic fluids. According to the dependence of
temperature to the energy of the particles, the temperature of
relativistic fluids such as photons will fall as $T\propto 1/a$.
For the non-relativistic fluids such as dust and cold dark matter,
temperature relaxes with a faster rate of $T\propto 1/a^2$. This
approach showed that the cooling of the cosmic fluid can not be
explained with the classical concepts from the thermodynamics and
cooling directly results from the expansion of universe which
makes the thermal peculiar velocity of particles reduce during the
stretch of the scales.


\end{document}